\documentclass[epj]{svjour}
\usepackage{amsmath,array,color,enumerate,epsfig,float,graphicx,latexsym,rotating,psfrag}
\begin{document}
\title{Effect of boundary scattering on spin-hall effect}
\author{S. K. Pandey\inst{1} \and T. P. Pareek\inst{2,3}
}                     
\offprints{shishir.kr.pandey@gmail.com}          
\institute{Department of Physics, National Taiwan University, No. 1, Sec. 4, Roosevelt Rd., Taipei-10617, Taiwan. \and 
Harish-Chandra Research Institute, Chhatnag Road, Jhunsi, Allahabad-211019 India. \and
Homi Bhaha National Institute, Anushaktinagar Mumbai, Maharashtra, India-400094.}
\date{Received: date / Revised version: date}
%
\abstract{
The spin dependent reflection in quasi-two-dimensional electron gas
from an impenetrable barrier in presence of Rashba and Dresselhaus 
spin-orbit coupling is analysed in detail. It is shown that the due to spin-orbit effects the 
 reflected beam split in two beams gives rise to multiple reflection
  analogous to phenomenon birefringence. The interplay between 
Rashba and Dresselhaus spin orbit coupling gives rise to anisotropy
 in Fermi energy surface and a non-zero net spin-polarized current 
 oscillating with two frequencies for all the values of incident 
 angle except at $45^{o}$ when averaged over all components of 
 reflected beam. It is also shown that in over critical region,all the three 
 polarization components as well as net polarization has non-zero values
 and are exponentially decaying as distance from the barrier increases
 which in turns spin-accumulation near the barrier is an important 
 consequence of spin-hall effect.
} 
\maketitle
\section{Introduction}
\label{intro}

The manipulation and coherent control of electronic spin degree of freedom 
has emerged as an important area of research in recent years. It in turn 
requires the ability to generate, inject and control spin polarized charge
current - an example of that is Datta-Das spin-transistor \cite{Datta}, where
a semiconductor is sandwiched between two Ferromagnetic contacts. The 
injection and detection of spin polarized current is achieved by using 
Ferromagnetic contacts in which spin is easier to manipulate because it 
behaves as a collective degree of freedom. In the semiconductor region, 
coherent control of spin polarized current is done using the band structure
spin-orbit(SO) coupling, known as Rashba SO interaction which arises due 
to structural inversion asymmetry\cite{Rashba}.

In recent years it has been realized that SO coupling can be used to efficiently
generate and detect spin-polarized current in semiconductor heterostructure 
without the Ferromagnetic contacts\cite{pareek}. In finite size
sample with SO coupling when an unpolarized charge currents passes it generates
spin currents(via SO scattering) in transverse direction which in turn leads 
to spin accumulation at the lateral edges of the sample and is known as Spin
-Hall effect(SHE)\cite{Kato,Sih,Wunderlich}. The Spin accumulation due to SHE has been observed experimentally\cite{Dyakonov,Hirsch,Zhang,Murakami,Sinova}. In such systems the spin accumulation 
and its relation to bulk spin current is a complex issue. 
This is because in finite size 
sample, the lateral edges of the sample acts like an impenetrable 
barrier for particles which carries spin. Therefore the boundary spin accumulation in such systems is affected by the elastic scattering of spin polarized carriers from the
sample boundary in presence of SO coupling. More precisely,
spin-dependent elastic reflection from an impenetrable barrier
in the presence of SOC depends on
the spin-orientation of particles which in-turn affects the spin accumulation. Hence spin accumulation at sample boundary not only depends
on the bulk spin current but also on the spin-dependent scattering from the lateral impenetrable barrier. For system with only Rashba SO coupling.
Spin-dependent elastic reflection of 2DEG  from an impenetrable barrier in the presence of
Rashba SOC was studied in Ref.\cite{winkler}, where it was shown to generate spin polarized reflected
beam.

The spin-orbit coupling owes its origin to appearance of inversion symmetry breaking electrical fields whether they arise intrinsically in the band structure(lack of inversion center) or by an external confining potential. In the former case inversion symmetry is broken locally and resulting SO interaction is known as Dresselhaus spin-orbit coupling\cite{Dress}, while in the later case
confining potential leads to structure inversion asymmetry which breaks inversion symmetry globally and leads to appearance of Rashba spin-orbit coupling\cite{Rashba}.
Beside these two, another 
type of SOC arises due to presence of heavy impurity known as Impurity induced
SOC. In low-dimensional nanosystem all these different type of SO couplings may be present simultaneously and compete with each other.
Among all these Rashba SOC is more important because of
its tunability via external gate voltage,while
Dresselhaus and impurity induced SOC is fixed and determined by the material properties,crystal structure and impurity type and concentration respectively.
Rashba and Dresselhaus SOC are also known as {\it intrinsic} SOC because of their origin in the band 
structure while impurity induced SOC is known as {\it extrinsic} SOC.
Although both Rashba and Dresselhaus SOC are linear in momentum, however, there is an important difference, i.e., 
the Rashba coupling is isotropic while Dresselhaus coupling is anisotropic, i.e., depends on the orientation of crystal. This is so because the Rashba coupling is determined by globally inversion asymmetry and is independent of crystal structure while Dresselhaus crucially depends on the crystal structure as it originates due to local inversion asymmetry\cite{}. The simultaneous presence of both Rashba and Dresselhaus couplings together with the tunability of Rashba SOC allows greater control over spin polarized transport which in turn gives rise to many interesting and novel phenomena such as, ballistic spin field effect transistor\cite{}, persistent spin helix and spin edge helices\cite{Badalyan} etc.

In view of this, we present a detailed study of 
spin dependent scattering in confined geometry
in presence of both Rashba an Dresselhaus SO coupling. The anisotropic nature of Dresselhaus SO coupling leads to an anisotropic Fermi energy. This anisotropy in conjugation with the tunability of Rashba coupling affects the spin dependent double refraction as well spin accumulation in a nontrivial way as 
we will see later. It is also shown that 
even if we take unpolarized incoming beam, the net polarization is coming out to
be non-zero. In the over critical region all the three component of spin polarization are 
present and are exponentially decaying as distance from the barrier increases. 

\section{Model and Anisotropic Fermi Surface}
\label{sec:1}
The model Hamiltonian including both Rashba and Dresselhaus coupling has the form,
\begin{equation}
H_{0}=\frac{\hbar^{2}k^{2}}{2 m^{*}}+\alpha(\sigma_{x}k_{y}-\sigma_{y}k_{x})+\beta(\sigma_{y}k_{y}-\sigma_{x}k_{x}),
\end{equation}
where $\mathbf{k}=(k_{x},k_{y},0)$ is the 2 in-plane wave vector and $m^{*}$ is effective mass.
The second and third term in E.(1) are Rashba and Dresselhaus SOC with $\alpha$ and $\beta$
as coupling coefficients respectively. The spin split eigen function and dispersion is 
\begin{equation}
\Psi_{\lambda}(\mathbf{r}) =\frac{e^{\mathbf{k \cdot r}}}{\sqrt{2}}
\begin{pmatrix}
e^{i\varphi_{\rho}} \\
\lambda
\end{pmatrix}
\end{equation}
\begin{eqnarray}
E_{\lambda}(\mathbf{k})=\frac{\hbar^{2}}{2 m^{*}}\left[ k^{2}+\lambda \frac{2k m^{*}}{\hbar^{2}}\lvert\rho(\alpha,\beta,\phi_{k})\rvert\right] \nonumber \\
\equiv \frac{\hbar^{2}}{2 m^{*}}\left[(k+\lambda \frac{m^{*}}{\hbar^{2}} \lvert{\rho(\alpha,\beta,\phi_{k})}\rvert)^2-\frac{{m^{*}}^{2}}{\hbar^{4}}\lvert{\rho(\alpha,\beta,\phi_{k})\rvert^{2}}\right]
\label{dispersion:1}
\end{eqnarray}
whith
\begin{equation}
\rho(\alpha,\beta,\phi_{k})=(i\alpha e^{-i\phi_{k}} -\beta e^{-i\phi_{k}}) \\
\lvert\rho(\alpha,\beta,\phi_{k})\rvert=\sqrt{\alpha^2+\beta^2-2\alpha \beta \sin(2\phi_{k})}
\end{equation}
where $\phi_{k}$ being the polar angle of in plane momentum $\mathbf{k} \equiv (k\cos{\phi_{k}},k\sin{\phi_{k}})$. The spinor phase
is given by $\varphi_{\rho}=Arg[\rho(\alpha,\beta,\phi_{k})]$ and $\lambda =\pm 1$ defines the
chirality of the eigenfunctions.
The Fermi wave vectors for a fixed energy($E_{F}$) are
\begin{equation}
k_{\lambda} = \frac{1}{\mu} \left(\sqrt{2\mu E_{F} + \vert \rho \vert^{2} } -\lambda \rho \right)
\end{equation}
where $E_{F}=E_{+}(k_{+})=E_{-}(k_{-})$. 

The spin split dispersion given by E.~(\ref{dispersion:1}) is shown in Fig.~(1) for various values of $\alpha$ and $\beta$
\begin{figure}
\centering
\includegraphics[height=3.3 cm,width=3.3 cm]{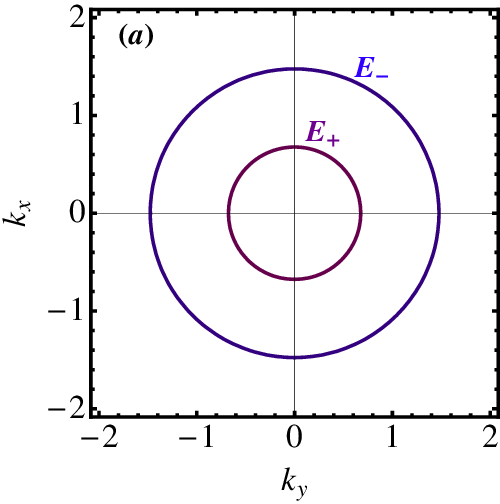}
\includegraphics[height=3.3 cm,width=3.3 cm]{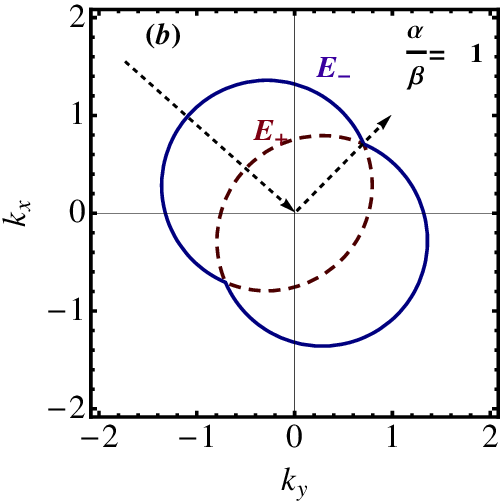}
\includegraphics[height=3.3cm,width=3.3cm]{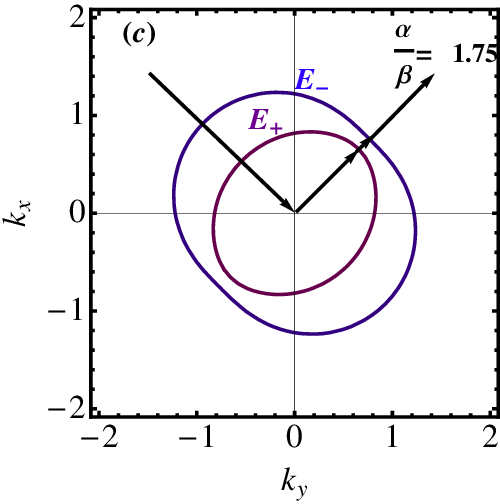}
\includegraphics[height=3.3cm,width=3.3cm]{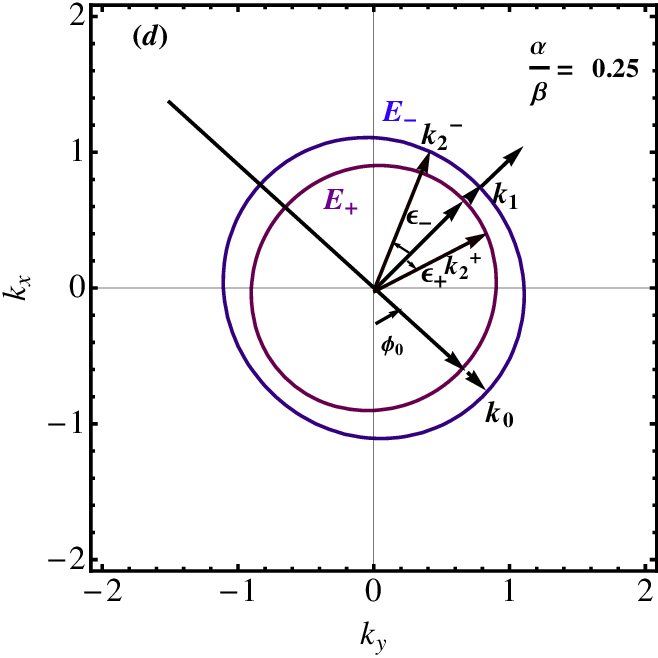}
\label{fig:1}
\caption{The figures(a-d) illustrating the dispersion curves for various values of $\alpha$ and $\beta$.
Fig.(a) corresponds to $\alpha= $ and $\beta=0.0$, for remaning figures(c-d) these parameters are shown in the figure itself. The two circles shows the constant energy contours in ($k_{x},k_{y}$) plane for the two chiral branches $E_{\lambda}$. Depending upon relative magnitude $\alpha/\beta$, these circles can cross each other (b) or become anisotropic as in (c).}
\end{figure}
In the dispersion curves shown above, the first corresponds(panel (a)) to the simple Rashba system($\beta$=0) and the rest three figures corresponds to the three different cases, namely for $\alpha = \beta$, $\alpha > \beta$, $\alpha < \beta$ respectively. 
The qualitative plot shows the asymmetric Fermi energy surfaces in the $k_{x}$-$k_{y}$ plane.
Note that when $\alpha$=$\beta$ (Fig.(b))
the curves touch each other along particular direction in k-space. 
This implies that for waves propagating along this direction spin splitting vanishes, and as a consequence precession induced spin dephasing 
ceases to act. This property was used in the non-ballistic spin field effect transistor proposed by
Loss\cite{}. For $\alpha\neq \beta$ (panel (c) and panel(d)), the two Fermi surface becomes anisotropic in  $k_{x}$-$k_{y}$ plane. Therefore simultaneous presence of both Rashba and Dresselhaus spin orbit coupling provides a much better control over the spin splitting and we will see later that this leads to interesting phenomena for spin dependent elastic reflection from impenetrable barriers.

\subsection{Elastic spin dependent scattering from impenetrable barrier} 
\label{sec:2}
We consider two dimensional system in xy plane with an impenetrable barrier V(x) along $\hat{x}$ axis,which is described by the Hamiltonian,
\begin{equation}
H=H_{0}+V(x),
\end{equation}
where $H_{0}$ is defined in E.1.
We assume that V(x)=0 for x$<$0 and V(x)=$\infty$ for x$>$0 and along $\hat{y}$ axis systems is free.
Consider an electron beam with chirality $\lambda$ and wave vector $k_{0}^{\lambda}$ ($\lambda$ incident on impenetrable barrier at an angle $\phi_{0}$. It is reflected elastically from the barrier and due to splitting of dispersion curves this processes generates two reflected waves, namely, ordinary reflected wave and extraordinary reflected wave. For ordinary reflection, energy as well momentum both are conserved while for extraordinary wave only energy is conserved. This is shown schematically in Fig.(2).
\begin{figure}[h]
\centering
\includegraphics[height=5.0cm,width=7.0cm]{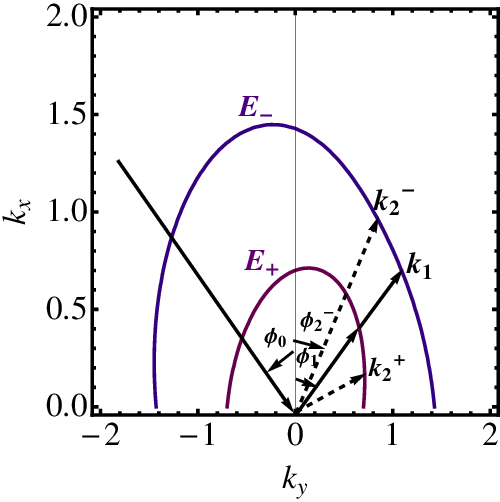}
\caption{Figure showing the ordinary and extra-ordinary reflection. The solid lines corresponds to ordinary reflected wave while dashed lines are used to show the extra-ordinary reflection. Here $\alpha/\beta=0.25$}.
\label{fig:2}
\end{figure}
The wave vectors for incident, ordinarily(momentum preserving) and extraordinary reflected waves are with $k_{in}^{\lambda}$, $k_{1}^{\lambda}$  and $k_{2}^{-\lambda}$ respectively. The most general scattering wave function for this system is the linear combination of these three and can be written as,following Ref.(\cite{}),
\begin{eqnarray}
\psi_k^{\lambda} (r)&=& \frac{\exp ^ {ik_{0 }^{ \lambda .r}} }{\sqrt{2A}\rho}\begin{pmatrix} i\alpha e^{-i\phi_{0}}-\beta e^{i\phi_{0}} \\
\lambda \rho \end{pmatrix} \nonumber \\ &+&\frac{A_{1}^{\lambda} \exp ^ {ik_1 ^ \lambda .r} }{\sqrt{2A}\rho}\begin{pmatrix}i\alpha e^{-i\phi _{1}^{\lambda}} -\beta e^{i\phi _{1}^{\lambda}} \\
\lambda \rho \end{pmatrix} \nonumber \\
&+&\frac{A_2^ {-\lambda} \exp ^ {ik_{2} ^ {-\lambda} .r} }{\sqrt{2A} \rho}\begin{pmatrix} i\alpha e^{-i\phi_{2}^{-\lambda}}-\beta e^{i\phi _{2}^{-\lambda}} \\ -\lambda \rho \end{pmatrix} \nonumber \nonumber
\end{eqnarray}
Translational invariance parallel to the barrier at $x=0$ implie that 
y-component of crystal momentum is conserved, hence,
\begin{equation}
k_{y} = k_{0}^{\lambda} \sin \phi_{0} = k_{1}^{\lambda} \sin \phi_{1} = k_{2}^{-\lambda} \sin \phi_{2}^{-\lambda}
\end{equation}
from which one obtains the angle of reflection for ordinary($\phi_{1}$) and extraordinary beams ($\phi_{2}$)as,
\begin{eqnarray}
\phi_{1} = \pi - \phi_{0} \\
\phi_{2}^{-\lambda} = \pi - \arcsin \left(\frac{k_{\lambda}}{k_{-\lambda}} \sin \phi_{0}\right)
\end{eqnarray}
In general $\phi_{1}$ and $\phi_{2}$ are different, implying that a single incident wave with a particularly chirality generate waves of both chirality as is shown in Fig.(2). The ordinary reflected wave is always propagating while the extraordinary wave may be propagating or evanescent depending on whether $\phi_{2}^{-\lambda}$ is real or imaginary. Since $k_ > k_{+}$
$\phi_{2}^{-}$ has always a real value for $0\leq \phi_{0}\leq \frac{\pi}{2}$ but in case of $\phi_{2}^{+}$ a real solution exits only 
for $\phi_{c}$ after which this angle becomes complex, where
\begin{equation}
\phi_{c} = \arcsin \left(\frac{k_{+}}{k_{-}}\right)
\end{equation}
The splitting angle i.e. the angular difference between the two reflected beam is 
\begin{equation}
\epsilon_{-\lambda} \equiv \phi_{0} - \arcsin \left(\frac{\sqrt{2\mu E + \vert \rho \vert^{2} } -\lambda \rho}{\sqrt{2\mu E + \vert \rho \vert^{2} }+ \lambda \rho} \sin \phi_{0}\right)
\end{equation}
The angle $\epsilon_{-}$ and $\epsilon_{-}$ have their largest value given by
\begin{eqnarray}
\vert \epsilon_{max} \vert & = & \frac{\pi}{2} - \phi_{c}
\end{eqnarray}
obtained at $\frac{\pi}{2}$ and $\phi_{c}$ respectively. The splitting angle $\epsilon_{-}$ is positive while $\epsilon_{+}$ is negative
From the conservation of $k_{y}^{\lambda}$, it can be seen that the $k_{x1}^{\lambda}$ and $k_{x2}^{-\lambda}$ become functions of $\phi_{1}$ 
and $\phi_{2}^{-\lambda}$ respectively. When $\phi_{0} > \phi_{c}$, $k_{2}^{+}$ becomes imaginary and leads to exponentially 
decaying current.
At the interface, the wave function must be continuous which yields the conditions 
\begin{equation}
{A_{1}^{\lambda}} = \left( \frac{e^{-2i\phi_{0}}-e^{-i\epsilon_{-\lambda}}}{1+e^{-i\epsilon_{-\lambda}}} \right ) \left\{ \frac{i\alpha e^{-i\epsilon_{-\lambda}}+\beta} {i\alpha e^{-i\epsilon_{-\lambda}}-\beta e^{-2i\phi_{0}}} \right\}
\label{a:1}
\end{equation}

\begin{equation}
A_{2}^{-\lambda} = \left(\frac{1+e^{-2i\phi_{0}}}{1+e^{-i\epsilon_{-\lambda}}}\right)\left\{\frac{i\alpha e^{2i\phi_{0}}-\beta e^{2i\phi_{0}}}{i\alpha e^{2i\phi_{0}}-\beta e^{i\epsilon_{-\lambda}}}\right\}.
\label{a:2}
\end{equation}
Using the above equation it is straight forward to obtain ordinary ($R_{\lambda \lambda}=\frac{A_{1}^{\lambda}}{A_{0}} $) and extraordinary ($R_{\lambda -\lambda}=\frac{A_{1}^{\lambda}}{A_{0}} $) reflection coefficients, 
\begin{equation}
R_{\lambda \lambda} = \left(\frac{\sin^{2} \left(\frac{\epsilon_{-\lambda}}{2}-\phi_{0}\right)}{\cos^{2} \frac{\epsilon_{-\lambda}}{2}}\right) \left\{\frac{\alpha^{2}+\beta^{2}+2\alpha \beta \sin \epsilon_{-\lambda}}{\alpha^{2}+\beta^{2}-2\alpha \beta \sin \left(\epsilon_{-\lambda} - 2\phi_{0}\right)}\right\} 
\label{r:1}\\ 
\end{equation}

\begin{equation}
R_{\lambda -\lambda} = \left(\frac{\cos^{2} \phi_{0}}{\cos^{2} \frac{\epsilon_{-\lambda}}{2}}\right) \left\{\frac{\alpha^{2}+\beta^{2}}{\alpha^{2}+\beta^{2}-2\alpha \beta \sin \left(\epsilon_{-\lambda} - 2\phi_{0}\right)}\right\}.
\label{r:2}
\end{equation}
In the above expressions (\ref{a:1},\ref{a:2},\ref{r:1},\ref{r:2}), the terms in curly brackets
reduces to one for $\beta=0$ which agrees with the previous result of Ref.\cite{}. In general the dependence on $\alpha$,$\beta$ is more complicated as is clear from the above expressions.To obtain insight and compare it with the simple Rashba system we plot reflection coefficient as a function of incident angles in Fig.~(3) for the three different cases i.e. $\alpha > \beta$,$\alpha < \beta $ $\alpha = \beta$.
\begin{figure}[h]
 \centering
  \includegraphics[height=3.4cm,width=5.2cm]{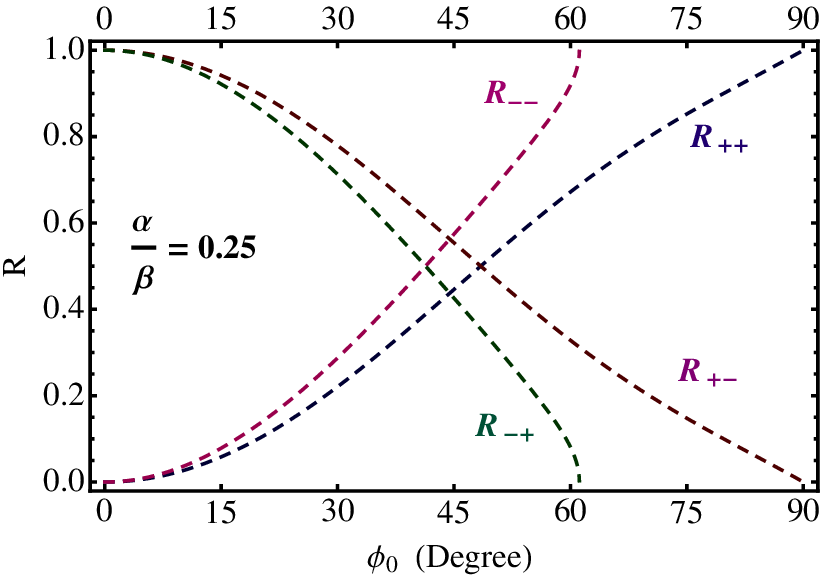}
  \includegraphics[height=3.4cm,width=5.2cm]{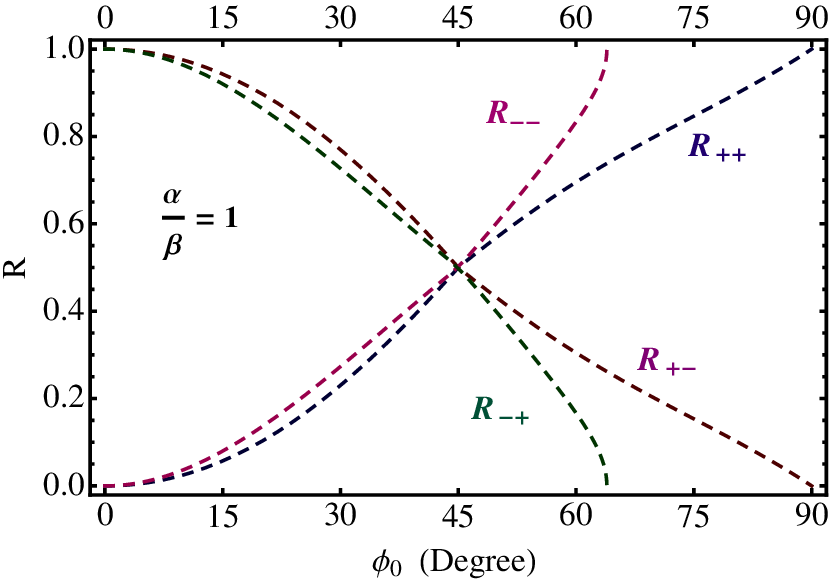}
  \includegraphics[height=3.4cm,width=5.2cm]{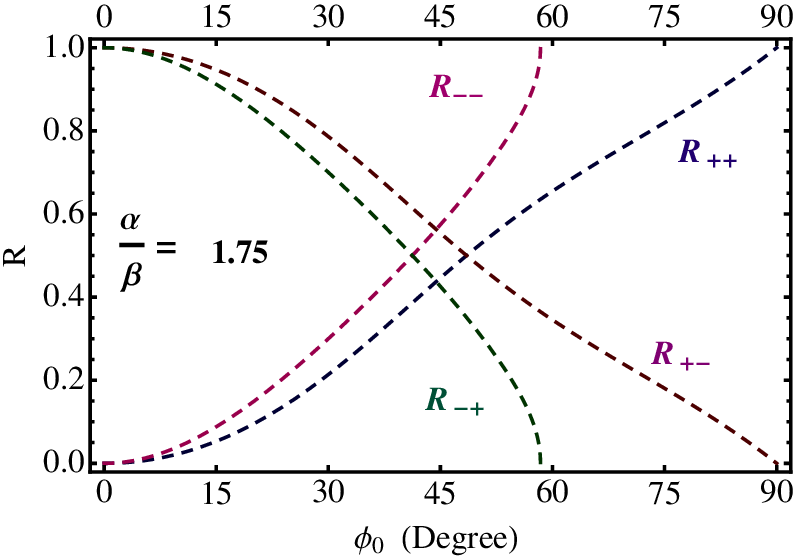}
  \caption{Reflection coefficient $R_{\lambda \lambda}$ for the three different ($\alpha > \beta$, $\alpha = \beta$,$\alpha < \beta$) cases for R and D 
  system as a function of incident angle is shown.}
  \label{fig:3}
\end{figure}
From Fig.(3) we notice that when the strength of Rashba and Dresselhaus SOC are different, the four reflection coefficients are never equal at any incident angle. This is consistent with the anisotropic Fermi contours 
in Fig(1) (panel(c) and (panel(d)). For $\alpha=\beta$ at 45 degrees all four reflection coefficients become equal since at this point the two Fermi contours cross each other. The is again reflected in the splitting angle which is plotted in Fig.(4), again the splitting vanishes for $\alpha=\beta$ at 45 degrees and for all other cases it never vanishes. We stress that this vanishing of
splitting angle or crossing of Fermi contours happens only if both Rashba and Dresselhaus couplings are present and of equal strength. If only Rashba or only Dressehaus coupling is present this is not so. 
In fact this is related to the fact that simultaneous presence of Rashba and Dressehaus coupling of equal strength introduces a
This implies that at a particular angle of incident the reflection from impenetrable barrier would not produce spin polarization only for $\alpha=\beta$ case and in this atypical case only the boundary spin accumulation will be fully determined by the bulk spin currents. However in quasi one dimensional systems (finite transverse width) since the electrons will be approaching boundary from all possible incident angles, therefore the boundary spin accumulation in general will be determined by both bulk spin current as well by the reflection from the boundary.

\begin{figure}[h]
 \centering
    \includegraphics[height=5.0cm,width=7.0cm]{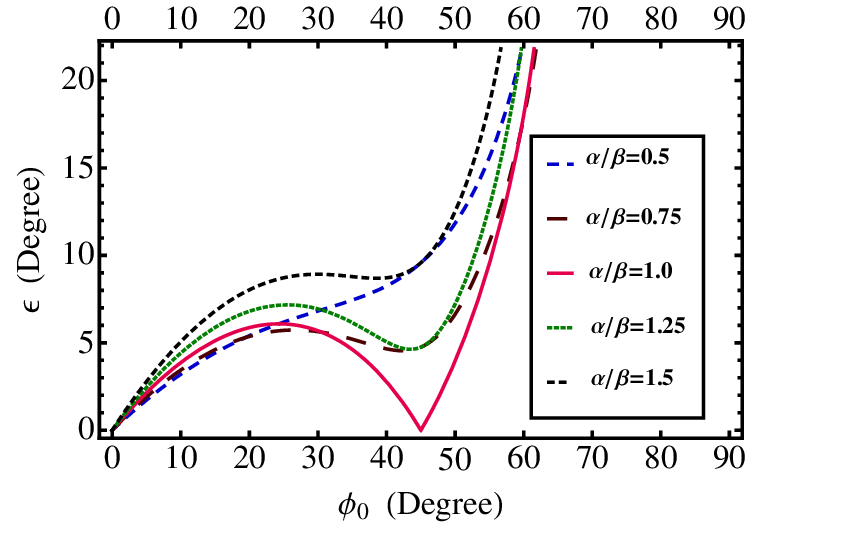}
  \caption{Total splitting angle as a function of incident angle is shown. It is shown that at $45^{o}$ when $\alpha = \beta$, 
  then the polarization is completely vanishes.}
  \label{fig:4}
\end{figure}

\section{Velocity and current}
We calculate the expression for the velocity Operator from the Hamiltonian (1) and is given by :
\begin{eqnarray}
{\bf v} &=& \frac{1}{\hbar} \left[\mu k + \alpha \left(\hat{j} \sigma_{x} - \hat{i} \sigma_{y}\right) - \beta \left( \hat{i} \sigma_{x} - \hat{j} \sigma_{y}\right) \right]
\end{eqnarray}

While considering the real angles $\phi$ the magnitude \textit{v} of the velocity :
\begin{eqnarray}
v \equiv \left\langle v \right\rangle &=& \frac{1}{\hbar}\left[\mu k_{\lambda}+ \lambda \left(\alpha - \beta \right)\right]
\end{eqnarray}
We see here that for real angles the velocity is same for all the beams but the velocity will be slightly higher for the complex angle
$\phi_{2}^{+}$.We do not here give the lengthy expression.
Also the probability current calculation gives :
\begin{eqnarray}
{\bf j} \equiv  \left\langle j \right\rangle &=& \frac{i}{\hbar} \left[\mu {\bf Re}\left(\left\langle \psi| k |\psi \right\rangle \right) + \alpha \left(\left\langle \psi | \hat{j} \sigma_{x} - \hat{i} \sigma_{y}| \psi \right\rangle \right)\right] \nonumber \\
&-&\frac{i}{\hbar} \left[\beta \left(\left\langle \psi | \hat{i} \sigma_{x} - \hat{j} \sigma_{y} | \psi \right\rangle \right) \right]
\end{eqnarray}
As expected, $j_{x}=0$ for an impenetrable barrier in both the cases of incoming and reflected beam while $j_{y}$, for the region 
in which both the components of beam are present, oscillates as a function of distance from the barrier.It can be realized by considering 
the interference terms of the three components of the wave function.
For the real angles, we get 
\begin{eqnarray}
J_{0}^{\lambda} &=& \vert A_{0} \vert^{2} v_{0} \\ 
J_{1}^{\lambda} &=& \vert A_{0} \vert^{2} \left(\frac{\sin^{2}\left(\frac{\epsilon_{-\lambda}}{2}-\phi_{0}\right)}{\cos^{2} \frac{\epsilon_{-\lambda}}{2}}\right) \nonumber \\
&\times &\left(\frac{\alpha^{2}+\beta^{2}+2\alpha \beta \sin \epsilon_{-\lambda}}{\alpha^{2}+\beta^{2}-2\alpha \beta \sin \left(\epsilon_{-\lambda} - 2\phi_{0}\right)}\right) {\bf v_{1}} \\
 J_{2}^{-\lambda} &=& \vert A_{0} \vert^{2} \left(\frac{\cos^{2} \phi_{0}}{\cos^{2} \frac{\epsilon_{-\lambda}}{2}}\right) \nonumber \\ &\times & \left(\frac{\alpha^{2}+\beta^{2}}{\alpha^{2}+\beta^{2}-2\alpha \beta \sin \left(\epsilon_{-\lambda} - 2\phi_{0}\right)}\right) v_{2}^{-\lambda}
\end{eqnarray}
In over-critical region i.e. for the complex angle $\phi_{2}^{+}$ :
\begin{eqnarray}
J_{2}^{+} &=& \vert A_{0} \vert^{2} \left(\frac{2\cos^{2} \phi_{0}}{\cos \epsilon_{max}}\right) \nonumber \\ 
&\times & \left(\frac{\alpha^{2}+\beta^{2}}{\alpha^{2}+\beta^{2}-2\alpha \beta \sin \left(\epsilon_{-\lambda} - 2\phi_{0}\right)}\right) v_{2}^{-\lambda} e^{2k_{2}^{+}x} {\bf v}_{2}^{+}
\end{eqnarray}
the current decays exponentially as the distance $|x|$ increases from the barrier. 

\begin{figure}
\centering
\includegraphics[height=4.2cm,width=7.0cm]{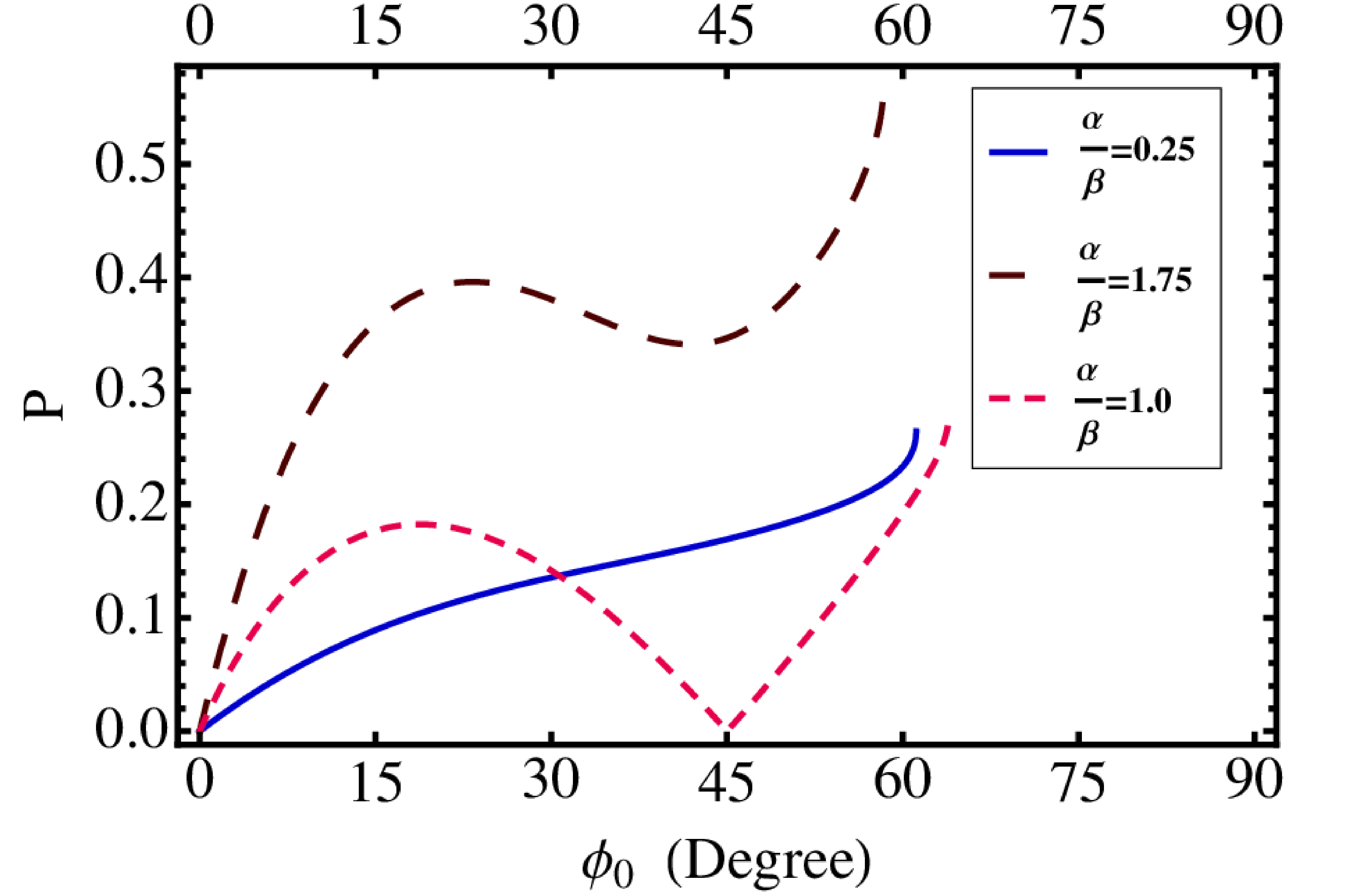}
\caption{Plot shows the net polarization of reflected beams against the incident angle at different values of 
$\alpha$ and $\beta$. Here interference terms are 
absent.}
\label{fig:5}
\end{figure}
Figure (5) shows the net polarization (which is the addition of polarization of ordinary and extra-ordinary reflected beams) at different Rashba and Dresselhaus SOC strength, against the incident angle. The plot at different SOC strength is shown from which it is clear that
net Polarization is non-zero at all the values of incident angle except at 45 degrees. 

\begin{figure}[h]
  \centering
  \includegraphics[height=4.2cm,width=7.0cm]{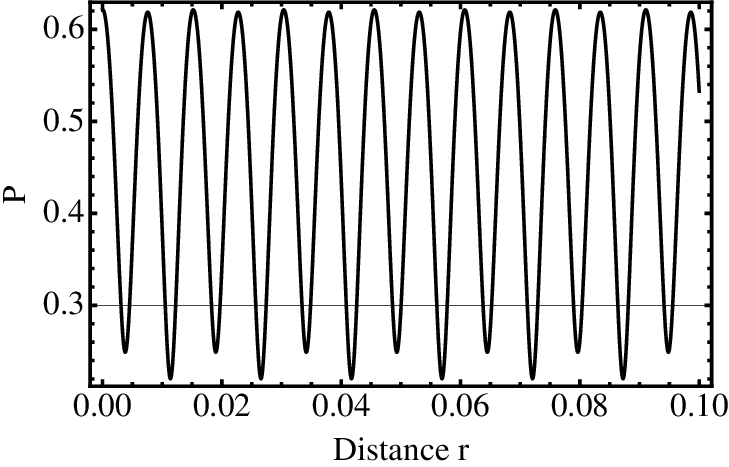}
  \caption{Plot for the net polarization of the beam incident at the angle $\frac{\pi}{3}$. Here the interference terms are 
  taken into consideration.}
  \label{fig:6}
  \end{figure}
 
We calculate the net polarization which is plotted against the distance r from the barrier which is shown in figure (6).
In this plot, value of Rashba SOC strength $\alpha$ is taken to be equal to the Dresselhaus SOC strength $\beta$. The contribution to this polarization is because of the two terms. In the first is term we have individually calculated and then added up the polarization for the incident, ordinarily and extraordinarily reflected beam. In the second term, polarization is calculated for the interference terms of these three components of the beam.The value of net polarization is always non-zero at all the values of the incident angle except at 45 degrees and oscillates with two different frequencies.

 \begin{figure}[h]
 \centering
  \includegraphics[height=3.8cm,width=5.8cm]{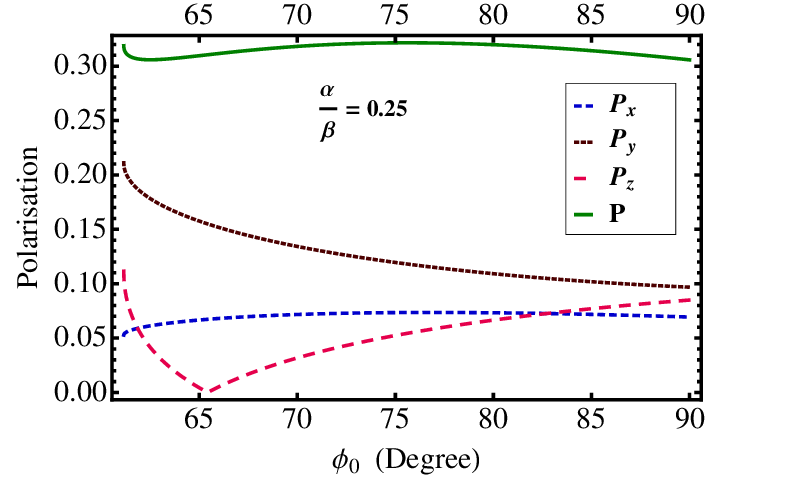}
  \includegraphics[height=3.8cm,width=5.8cm]{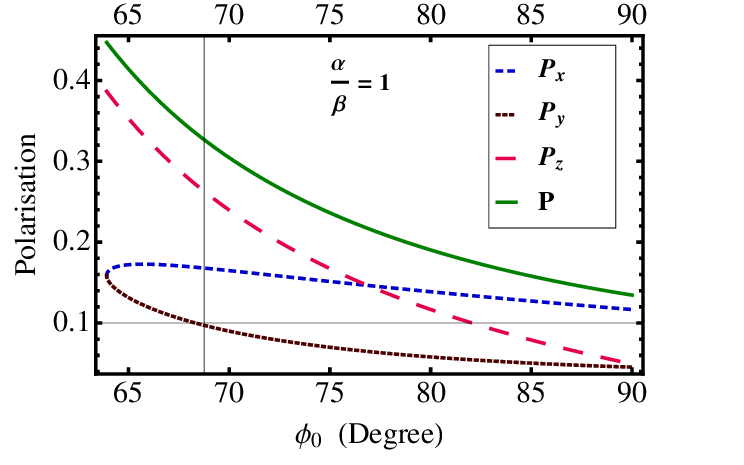}
  \includegraphics[height=3.8cm,width=5.8cm]{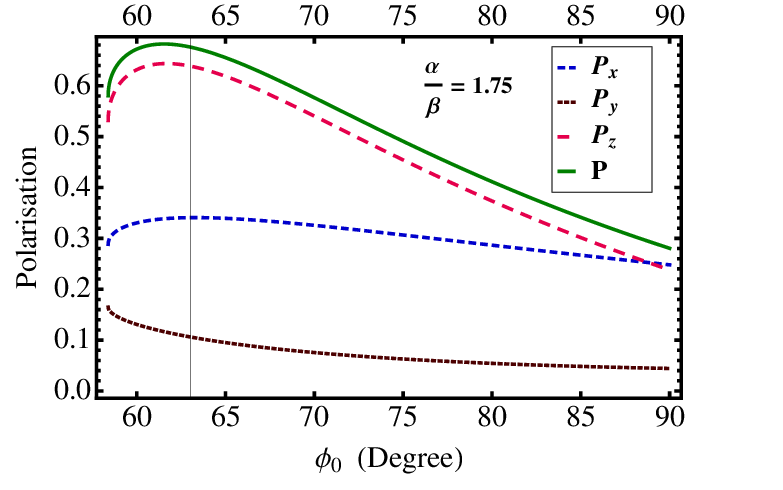}
  \caption{Different components of polarization with non-zero net polarization of extraordinary reflected beam in 
  over critical region is plotted against incident angle $\phi_{0}$ for the 
  different values of $\alpha$ and $\beta$. $P_{x},P_{y},P_{z} $ are the x,y and z component of polarization. P is the net polarization. }
  \label{fig:7}
\end{figure}
In the figure(7), for the over-critical region it is clearly shown that all the three components of polarization of extraordinary reflected 
beam will be present irrespective of the three cases of SOC strengths $\i.e. $ $\alpha < \beta$,$\alpha > \beta$ and $\alpha = \beta$.
Also in this region the net polarization always has non-zero values.
All the three components  of polarization as well as the net polarization will exponentially decay as the distance from 

\begin{figure}[h]
  \centering
  \includegraphics[height=4.2cm,width=7.0cm]{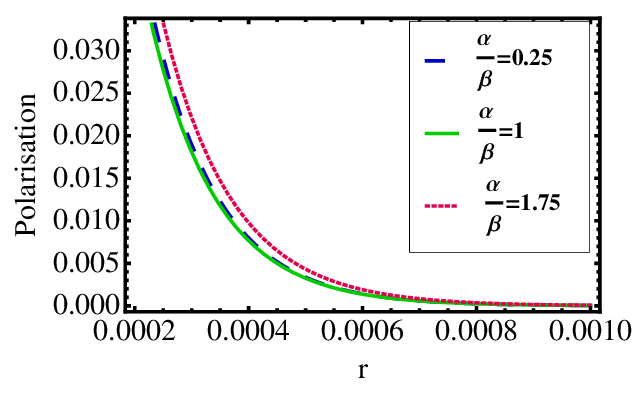}
  \caption{Plot for the net polarization of the beam incident at the angle $\frac{\pi}{3}$. Here the interference terms are 
  taken into consideration.}
  \label{fig:6}
  \end{figure}

the barrier increases. This is shown in figure(8). 
This non-propagating non-zero polarization gives rise to spin-accumulation near the barrier
which is clearly an important consequence of spin-dependent elastic reflection from a impenetrable barrier in the presence of R and D 
SOC. So it is clear that spin-accumulation near the barrier is a typical phenomena in which spin-accumulation near the barrier  
will not only depend upon the bulk spin-current but also on the spin-dependent scattering from the lateral 
impenetrable barrier which can be think of the lateral edges of the sample.

\section{Conclusion}
In conclusion, present work shows that the spin-orientation 
of the electrons changes because of the spin-dependent reflection from
an impenetrable barrier in the presence Rashba and Dresselhaus SOC. 
It gives rise to a new mechanism of multiple reflection 
analogous to the birefringence phenomena. Also an important feature of 
anisotropy in the Fermi energy surface come into appearance because 
of this reflection. We also observed that increase in the value of 
Dresselhaus SOC strength $\beta$ gives increases the 
anisotropy in the system. In this case a non-zero spin-polarized 
current is observed for all incident angle between
0 to $\frac{\pi}{2}$ except at angle of $45^{o}$ which is clearly 
an another important feature of Rashba and Dresselhaus SOC interplay.
Also it is shown that in the over-critical region,there is some spin-accumulation
 near the barrier is also affected by scattering of the bulk spin-current from the lateral edges of 
 the sample which in turns affect the spin-accumulation as in the case of spin-hall effect.

\end{document}